**Skopje and Sofia 2005 Earthquake and Geomagnetic data and the Geomagnetic Quake as Imminent Reliable Earthquake's Precursor**

**S. Cht. Mavrodiev[1] and L. Pekevski[2]**

[1]*Institute for Nuclear Research and Nuclear Energy, Bulgarian Academy of Sciences, Sofia, Bulgaria*

[2] *Seismological Observatory, Faculty of Natural Sciences and Mathematics, Skopje, Macedonia*

Correspondence to: S.Cht.Mavrodiev (mavrodi@inrne.bas.bg)

**Abstract**.

The imminent "when" earthquake's predictions are based on the correlation between geomagnetic quakes and the incoming minimum (or maximum) of tidal gravitational potential. The probability time window for the incoming earthquake is for the tidal minimum approximately ± 1 day and for the maximum- ± 2 days. The statistic evidence for reliability is based on of distributions of the time difference between occurred and predicted earthquakes for the period 2002- 2005 for Sofia region and 2004- 2005 for Skopje.

The project for complex Balkan- Black Sea region NETWORK for earthquake prediction by using the reliable precursors will be proposed in near future. The Project is based on the temporary data acquisition system for preliminary archiving, testing, visualizing and analyzing of the data with aim to prepare regional daily risk estimation.

## 1   Introduction

The problem of "when, where and how" earthquake prediction can't be solved only on the basis of seismic and geodetic data (Aki, 1995; Pakiser and Shedlock, 1995; Geller et al., 1997, Main, 1999a,b; Ludwin, 2001).

The possible tidal trigging of the earthquakes has been investigated for a long period of time (Knopoff, 1964; Tamrazyan, 1967; Tamrazyan, 1968; Ryabl at al., 1968; Shlien, 1972; Molher, 1980; Sounau et al., 1982; Shirley, 1988; Bragin, 1999). The conclusion that the earthquake's time is correlated with the tidal extremums is not unique, because in some of the extremums there are not earthquakes.

The including of additional information in the monitoring, for example, the analysis of the Earth electrical currents signals, permits to estimate the most probable time of incoming earthquake (Thanassoulas, 1991, Thanassoulas et al., 2001a, b).

The more accurate space and time measuring set for the Earth's crust condition parameters, the including in the monitoring of the electromagnetic fields measurements under, on and over the Earth surface, the temperature distribution and other possible precursors can be useful for the study of the "when, where and how" earthquake's prediction. For example in the papers (Varotsos and Alexopoulos, 1984a, b; Varotsos at al., 1996; Geller, R. J., 1996), the possibility for short-term earthquake prediction in Greece by seismic electric signals was investigated and analyzed. The results of observations on seismo-electromagnetic waves at two earthquake experimental areas in China were presented in the paper of Qian at al., 1994.

The atmospheric and ionospheric electromagnetic phenomena associated with earthquakes were analyzed in many books and papers (Hayakawa and Fujinawa, 1994; Hayakawa et al., 1999, Hayakawa et al., 2000, Hayakawa and Molchanov, 2002) and the future direction of investigation related to earthquake prediction was proposed, as well as its practical application to some concrete events. The papers (Oike and Ogawa, 1982, 1994) concern the observations of electromagnetic radiation in the LF and VLF ranges related with the occurrence of earthquake. The results of complex investigation of the variations of crust electrical resistivity as a function of tidal deformations on the basis of extremely low frequency radio station, which permit the hope for increase the reliability of electromagnetic based earthquake prediction, are presented in paper of Saraev et al., 2002.

Evidence in papers (Eftaxias, 2001, 2002) is presented that electromagnetic anomalyties in wide range of radio frequencies from ULF, VLF to VHF have been observed before some destructive earthquakes in continental Greece.

The impressive results of modified VAN method are presented in papers and web site (Thanassoulas, 1991, 1999; Thanassoulas et al., 2001 a, b, c, d), where, the appropriate measuring of electric Earth signals and their analysis demonstrates that the direction to the epicenter of incoming earthquake can be estimated and the time is defined from the next extremum of tidal potential. Some possible geophysical models of the phenomena are proposed and the prediction of the future magnitude is analyzed. The including of more than one site in the monitoring will permit short term earthquake prediction and will give some estimation for the magnitude (Thanassoulas, 1999).

In web site (ws) (Ustundag, 2001) and papers cited there, are presented the results of electropotential monitoring, based on the special constructed electrometer and appropriate temporal data acquisition system, for researching the electropotential variations as earthquake precursor.

One has to mention the satellite possibilities for monitoring the radiation activity of the Earth surface for discovering the radiation variations (ws Dean, 2003), temperature anomalies (Ramesh, 2004) and earthquake clouds (Zhonghao Shou, 1999, 2005) which can serve as earthquake precursors.

The analyses of the data from satellite monitoring for the ionosphere and the Earth radiation belt parameters also give evidences for anomalyties which can be interpreted as earthquake precursors. The information for the some results from the developing of earthquake precursor research could be found in the conference sites: Contadakis, 2002 and Papadopoulos, 2003.

The data for the connection between incoming earthquake and meteorology effects like quasistationary earthquakes clouds can be seen in the site of Zhonghao Shou, 1999. The statistic from 1993 for the reliability of prediction is also represented together with some theoretical models and estimations for the effect.

In order to summarize the results we can say, that the standard geodetic monitoring (USGS Pf, ws, 2002; Pakiser L, Shedlock K.M., 1995), the monitoring of different component of electromagnetic field under, on and over the Earth surface, some of the atmospheric anomalyties and behavior of charge distribution in the Earth radiation belts (see for example Silina, 2001; Larkina and Ruzhin, 2003), sometime could serve as unique earthquake's precursor. It is obvious, that for solving the reliability problem all different approaches should be unified, including the biological precursor data.

The progress in geomagnetic quake earthquake precursor approach (Mavrodiev, Thanassoulas, 2001) is presented (Mavrodiev, 2002 a, b, 2003 a, b, c; Mavrodiev, 2004). The approach is based on the understanding that earthquake processes have a complex origin. Without creating of adequate physical model of the Earth existence, the gravitational and electromagnetic interactions, which ensure the stability of the Sun system and its planets for a long time, the earthquake prediction problem can not be solved. The earthquake part of the model can be repeated in the infinity way "theory- experiment-theory" using nonlinear inverse problem methods looking for the correlations between fields in dynamically changed space and time scales. Of course, every approximate model (see for example Thanassoulas, 1991; Thanassoulas et al., 2001a, b) which has some experimental evidence has to be included in the analysis. It seems obvious that the problem of adequate physical understanding of the correlations between electromagnetic precursors, tidal extremums and incoming earthquake is connected with the progress of the adequate Earth's magnetism theory.

The achievement of the Earth's surface tidal potential modeling, which includes the ocean and atmosphere tidal influences, is an essential part of the research. In this sense the comparison of the Earth tides analysis programs (Dierks and Neumeyer, ws) for the ANALYZE from the ETERNA-package, version 3.30 (Wenzel, 1996 a, b), program BAYTAP-G in the version from 15.11.1999 (Tamura, 1991), Program VAV (version from April 2002) of Venedikov et al, 2001, 2003 is very useful.

The role of geomagnetic variations as precursor could be explained by the obvious hypothesis that during the time before the earthquakes, with the strain, deformation or displacement changes in the crust, in some interval of density changing arise the chemical phase shift which leads to an electrical charge shift. The preliminary Fourier analysis of geomagnetic field gives the time period of alteration in minute scale. Such specific geomagnetic variation we call geomagnetic quake. The linear piezo- effect model for electrical currents, which produce such geomagnetic variations, can not explain the alternations. The results from laboratory researching of electromagnetic effects in stress conditions can serve as qualitative support of the quantum mechanic phase shift explanation for mechanism generating the geomagnetic quake before earthquake and others electromagnetic phenomena in the time of earthquake (Freund et al, 2002; St-Laurent et al, 2006)

The K-index (Balch, 2003), accepted for the estimation of the geomagnetic conditions, can not indicate well the local geomagnetic variation for time minutes period, because it is calculated as a first moment of the geomagnetic filed on the basis of 3-h data. Nevertheless, the K- index behavior in the near space has to be analyzed, because of the possible Sun wind influence on the local behavior of the geomagnetic filed.

If the field components are measured many times per second, one can calculate the frequency dependence of full geomagnetic intensity and to analyze the frequency spectrum of geomagnetic quake. If the variations are bigger than usual for some period of time, one can say that we have the geomagnetic quake, which is the earthquake precursor. The nonlinear inverse problem analysis for 1999- 2001 of geomagnetic and earthquake data for Sofia region gives the estimation, that the probability time window for the predicted earthquake (event, events) is approximately ± 1 day for the minimum Earth the tidal potential and ± 2 days for the maximum.

The future epicenter coordinates could be estimated from the data from at least 3 points of measuring the geomagnetic vector, using the inverse problem methods, applied for the estimation the coordinates of the volume, where the phase shift arrived in the framework of its time window. For example the first work hypothesis can be that the main part of geomagnetic quake is generated from the vertical Earth Surface- Ionosphere electrical current.

In the case of incoming big earthquake (magnitude > 5 - 6) the changes of vertical electropotential distribution, the Earth's temperature, the infrared Earth's radiation, the behavior of water sources, its chemistry and radioactivity, the atmosphere conditions (earthquakes clouds, etc.) and the charge density of the Earth radiation belt, have to be dramatically changed near the epicenter area.

The achievements of tidal potential modeling of the Earth's surface, including ocean and atmosphere tidal influences, multi- component correlation analysis and nonlinear inverse problem methods in fluids dynamics and electrodynamics are crucial for every single step of the constructing of the mathematical and physical models.

In Sect 2 are given the statistics evidences for the reliability of the time window earthquake prediction on the basis of the geomagnetic field measurements (Mavrodiev, 2002b) and the Earth tidal behavior (Venedikov at al., 2002) for Sofia region (2002-2005) and for Skopje one (2004-2005)

## 2. The geomagnetic field quake as a reliable imminent time window earthquake's precursors for Balkan, Black Sea region- Sofia, Skopje data

As a support for complex approach it is useful to stress that the author's interests to the earthquake's prediction problem arise as a result of complex research of the Black Sea ecosystem about 15- 20 years ago (Mavrodiev, 1998). During the time of gathering the historical data for the ecosystem, was observed, that the Crime earthquake, occurred in 1928, is an evidence for electromagnetic and earthquake correlations. Such hypothesis has been proposed by the academician Popov in the early 30-ties of the 20 century (private communication).

According the INTERMAGNET (Geomagnetic data, 1986) requirements for measuring the geomagnetic field (see Fig. 1) on Earth's surface (http://www.intermagnet.org), the accuracy is ±10 nT for 95% of reported data and ± 5 nT for definitive data, with one sample per 5 seconds, in the case of vector magnetometer (F (XYZ) or F (HDZ)) and 1 nT, with 3 samples per second, for Scalar Magnetometer (F).

In Sofia, the geomagnetic vector projection $H$ is measured with relative accuracy less or equal to 1 nT by a fluxgate, feedback based device of rather original and simple, but powerful construction. (know-how of JINR, Dubna, Boris Vasiliev, 1998, private communication). It is used with 2.4 samples per second. Due to technical reasons the sensor was oriented under the Horizon in a manner that the measured value of $H$ is around 20 000 nT.

Analyzing the correlations between the behavior of the geomagnetic field, Earth tidal gravitational potential and the occurred earthquakes (from 1999 to 2001) it turns out that the daily averaged value of $\sigma_{Hm}$ and $\sigma_{\Delta Hm}$, which we denote by $Sig$ ( $Sig$), is playing the role of earthquake precursor.

Fig. 1 illustrates the behavior of geomagnetic field component $H_m$ and its variation $\sigma_{Hm}$ for a period without earthquakes precursor in the region. Figs 2 and 3 illustrate the behavior of geomagnetic field and its variation, which is unusual. Is this cases there are a geomagnetic quake, which is precursor for incoming event (earthquake or earthquakes)? One has to be sure that there are not a cosmos or Sun wind reasons for the geomagnetic quake (ws [NOAA, 1972].

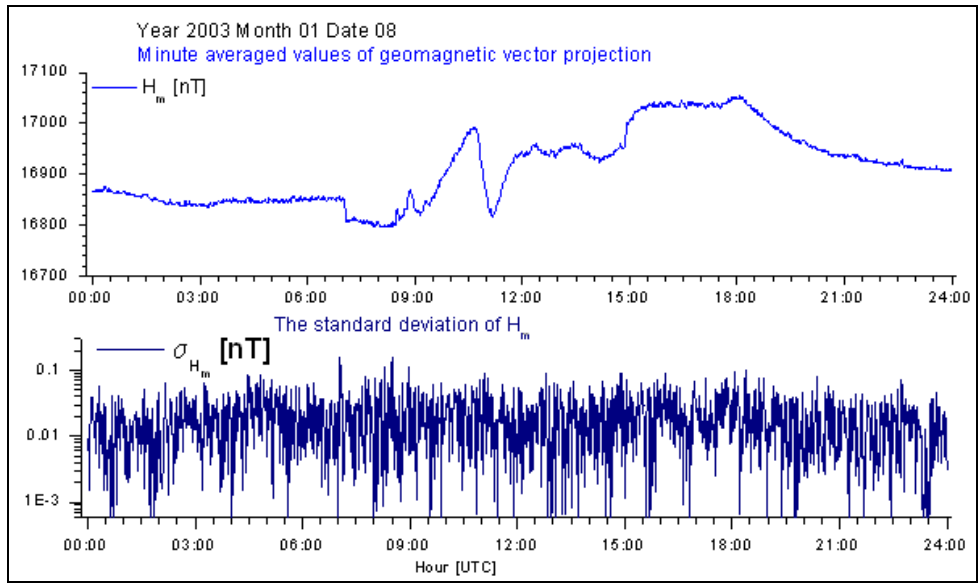

Fig.1. The behavior of geomagnetic field for a normal day.

For example, in Fig. 4 the predicted time window was 3 ± 1 June, 2002 and the prediction was confirmed with earthquake, occurred at 03/06/2002 02:04, Lat 41.95N, Lon 23.10E, h=8km, ML=2.6, 50 km from Sofia.

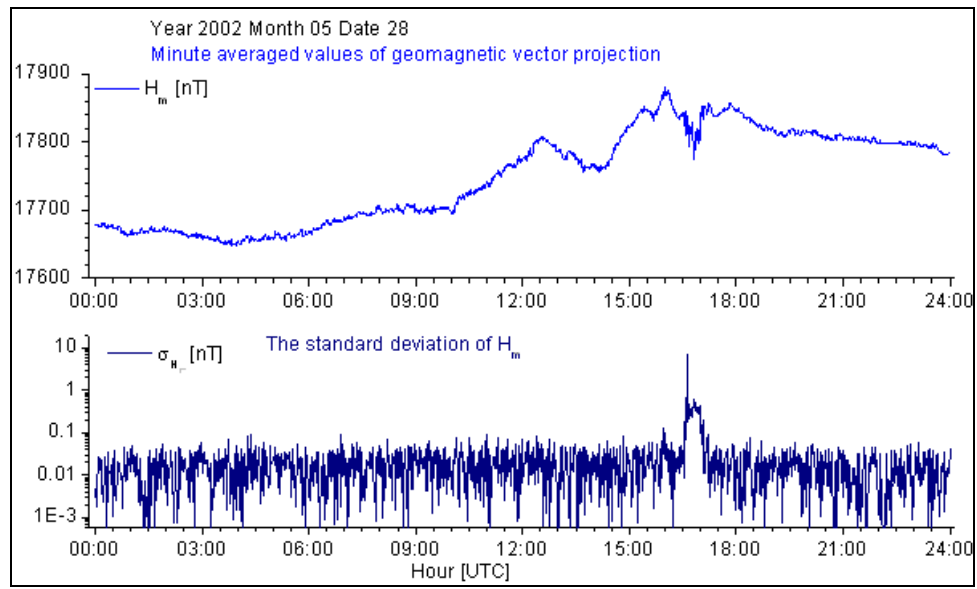

Fig.2. The behavior of geomagnetic field for a day with a signal for a near future earthquake.

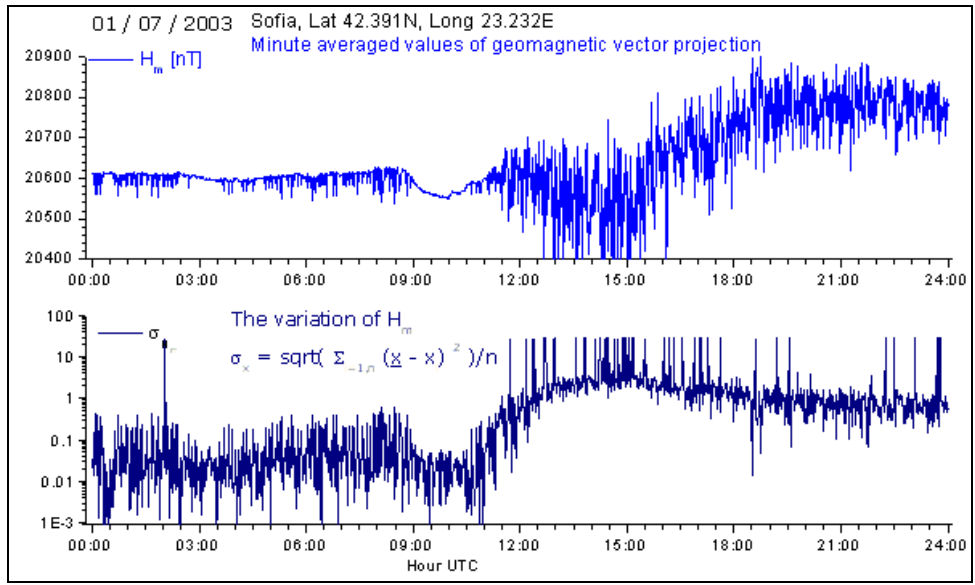

Fig.3. Day with geomagnetic quake, which is precursor for more than one earthquake.

The preliminary Fourier analysis of $H_m$ data gives the fact that the bigger geomagnetic variations are caused from the arriving for hours time period new frequencies with periods from 10-th of seconds till 10-th of minutes and with very specific amplitude behavior. Such spectrum, which arrives for hour period of time, is invisible for minute samples measuring. Its almost real time Fourier analysis is very complicated and there are some digital evidences, that the arrived electromagnetic field can not a linear piezo effect explanation.

The probability time window of the incoming event (or events) is defined by the next date of the Earth tidal potential extremum with tolerance approximately for the tidal minimum ± 1 day and for the maximum ± 2 days- Fig. 6.

The uncertainty of distinguishing the predicted event (or group of events – for example aftershocks) from the events which occurred in the region at different distances and magnitudes in the predicted time window is solved on the basis of inverse problem methods with new earthquake's influence characteristic function $S_{ChtM}$:

$$S_{ChtM} = 10^M/(Req+Distance)^2, Req=0.40+Depth \text{ [hundred km]} \quad (1)$$

The physical sense of the function $S_{ChtM}$ is that it is proportionally to the density distribution on the Earth surface of earthquake's energy. It is important to point out that the first consideration of the magnitude and distance dependences was obtained on the basis of nonlinear inverse problem methods. Obviously, the nearer and biggest earthquake (relatively biggest value of $S_{ChtM}$) will bear more electropotential variations, which will generate more power geomagnetic quake which depends on the depth and geology.

At this stage of the study, as a measure of daily geomagnetic state serve the value of averaged for 24 hours (1440 minutes) standard deviation $\sigma_{Hm}$, the signal function Sig and its error δSig are

$$\text{Sig} = \Sigma_{m=1,M} \, \sigma_{Hm}/M, \quad \delta\text{Sig} = \Sigma_{m=1,M} \, \delta\sigma_{Hm}/M, \qquad (2)$$

where $\sigma_{H\mu}$ is the standard deviation

$$\sigma_{H\mu} = \text{sqrt} \, \Sigma_{t=1,N} \, (H_t - H_\mu)^2/N, \quad \delta\sigma_{H\mu} = \text{sqrt} \, \Sigma_{t=1,N} \, (\delta H_t - \delta H_\mu)^2/N,$$

$H_m$ is minute average value of geomagnetic vector projection $H_i$, measured 2.4 times per second, N= 140, M= 1440:

$$H_\mu = \Sigma_{t=1,N} \, H_t/N, \quad \delta H_\mu = \Sigma_{t=1,N} \, \delta H_t/N,$$

the minutes per day are M=1440, and samples per minute are N=144.

The numerical definition of geomagnetic quake is the condition (3)

$$\text{Sig}_{\text{Today}} - \text{Sig}_{\text{Yesterday}} > |\delta\text{Sig}_{\text{Today}} + \delta\text{Sig}_{\text{Yesterday}}| / 2, \quad (3)$$

where δSig is the error of Sig.

If criteria (3) are fulfilled and there are not a Cosmos and Sun generated variations of the geomagnetic field one has to conclude that the geomagnetic quake is precursor for incoming earthquake. In the next minimum or maximum of the local tidal gravitational potential somewhere in the region this predicted earthquake will occur.

For some of the cases, the criteria (3) have to be calculated for decades of hours or ten's of minutes.

The signal for earthquakes with different epicenters we observe in the case when the specific behavior of field and its standard deviation occurs more than ones per day in different hours.

The analysis of the precursor function *Sig* on the basis of special digital 5 points derivatives can serve in the future for creating the algorithm for automated alert system.

It is next the more detailed time window can be achieved by analyzing the daily variations of tidal potential, calculated every 3 hours (see for example the definition of K- index).

At this stage of the study all earthquakes have the same $S_{ChtM}$ for different definitions of the magnitude. After the developing of mathematical models of empirical and theoretical dependences between incoming earthquake processes, geomagnetic quake and parameters of earthquake on the basis of inverse nonlinear problem we will obtain a set of $S_{ChtM}$ functions in correspondence with the different definition of magnitude and in correspondence with the earthquake's energy. The earthquake volume parameters, its depth, chemical and geological structures of the region have to be included in the dependences as well.

Fig. 4 illustrates the possibility for registration of big world earthquakes with magnitude greater than 6 by specific behavior of the geomagnetic change which has a place in about 60 % of the cases in monitoring period. The researching of this correlation will give information on the global Earth electric current distribution and its dependences on Sun wind variations.

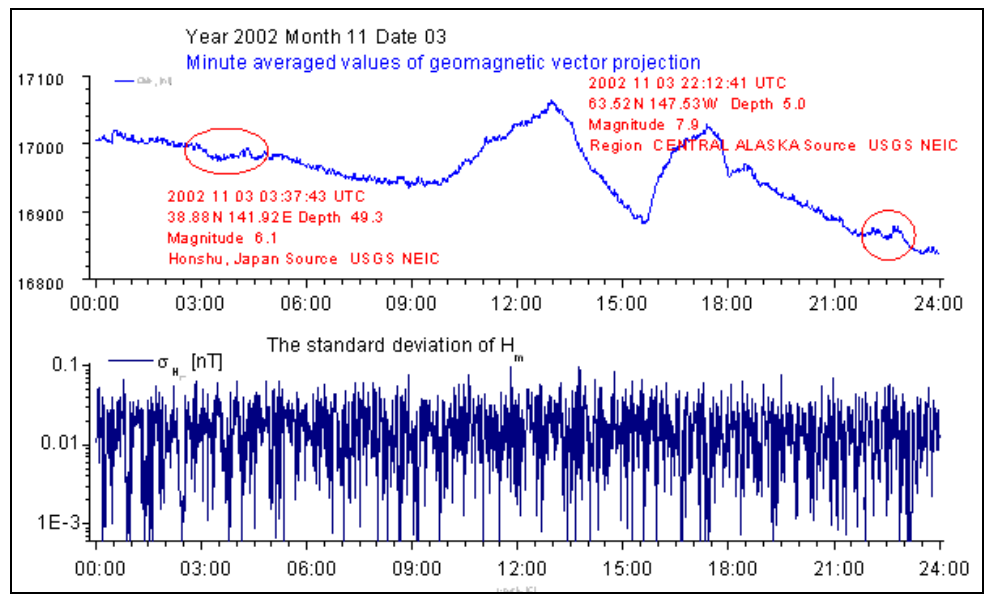

Fig.4. Registration of big world earthquake with Magnitude > 6.

## 2.1. Sofia data

In the next Fig.5 up to down the daily and hour behavior of the tidal potential, earthquake's energy density $S_{ChtM}$, magnitude, distance from Sofia (vertical lines from the magnitude) and function *Sig* are presented as they are published in the directory "Every day monitoring" of the website "Earthquake prediction using reliable earthquake precursors-http://theo.inrne.bas.bg/~mavrodi/" for the period June- October, 2005. It is seen that after fulfilling the condition (3) in the next tidal extremum there is occurring event in the region, which is identifying by the maximum of function (1).

For convenience to test the predictions in the directory "Monthly monitoring" are presented the functions Hm, its daily averaged and Sig as in Fig.

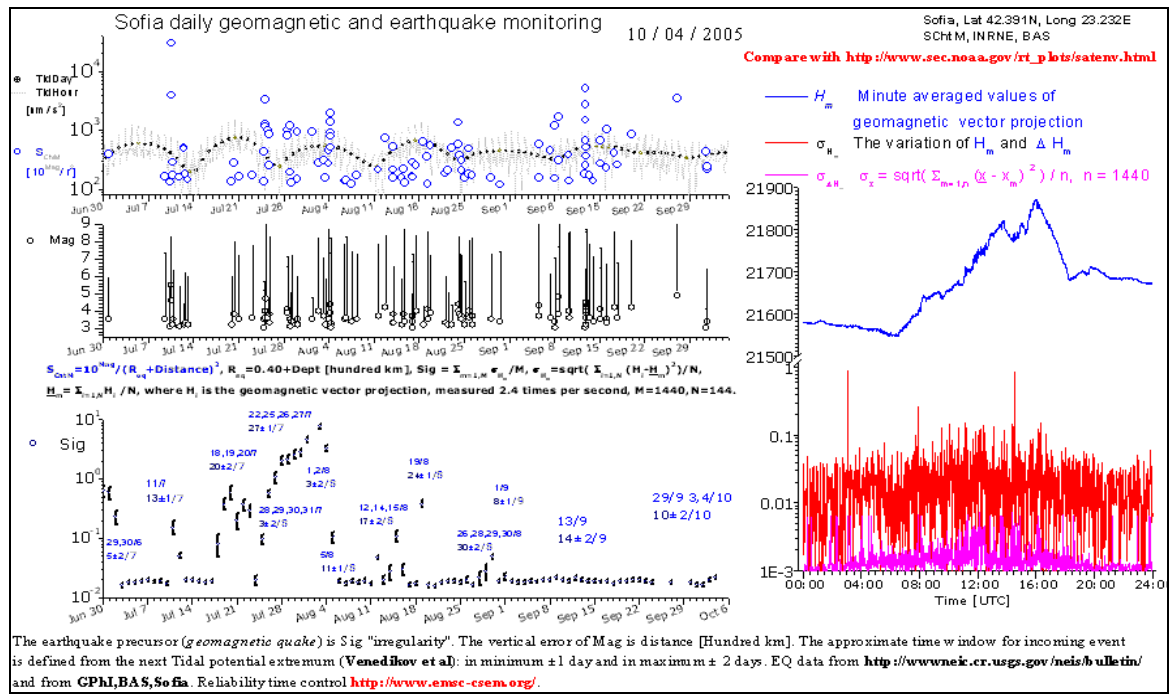

*Fig.5. The reliability of the time window prediction for incoming earthquake- June- October, 2005, Sofia region.*

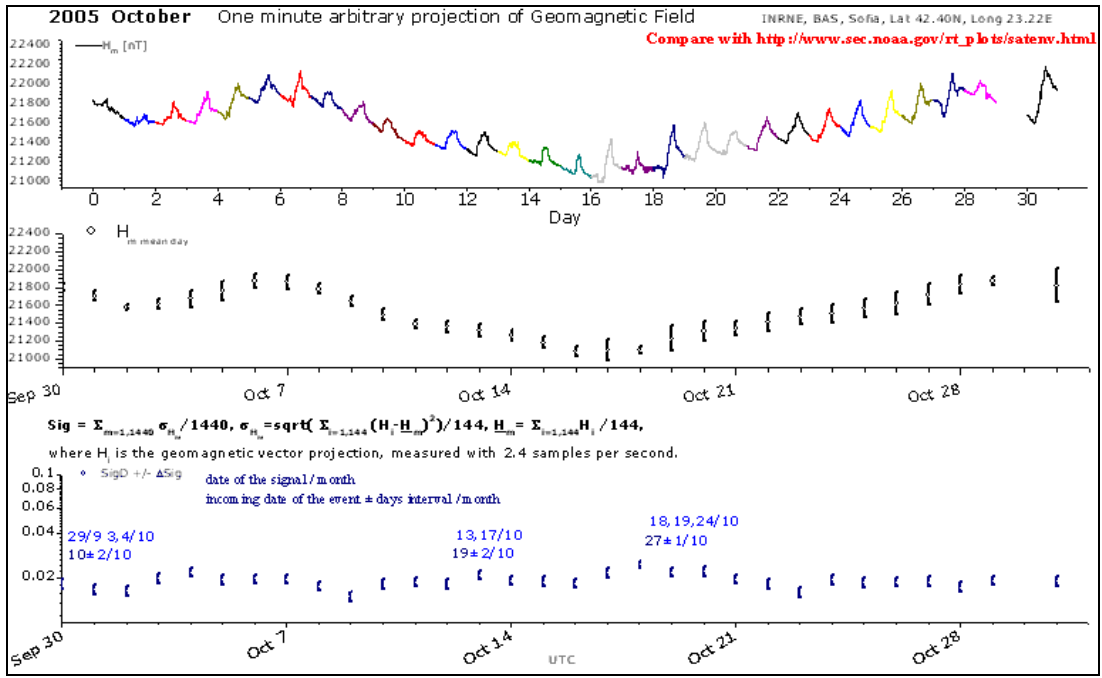

*Fig.6. The "monthly monitoring" figure- October, 2005, Sofia region.*

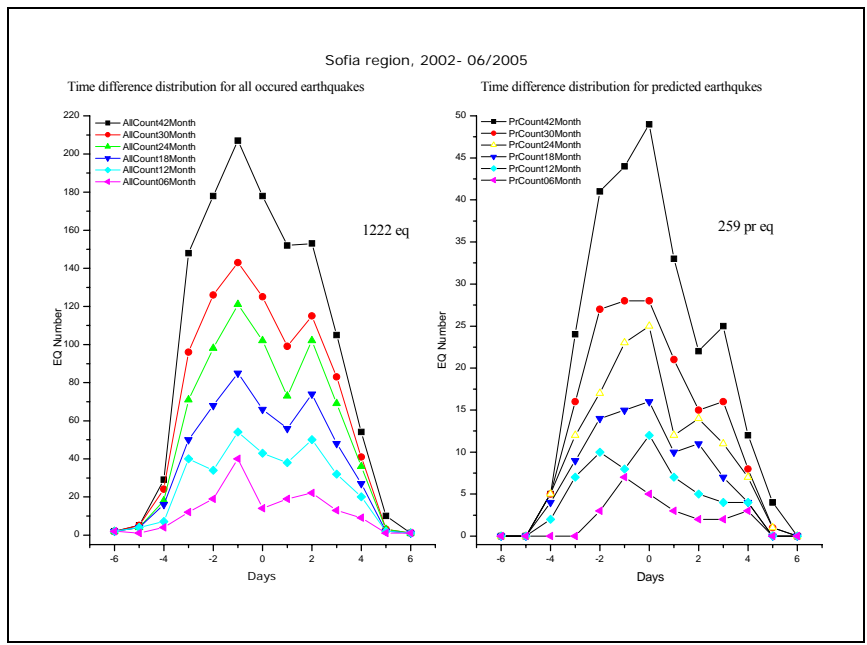

*Fig7. The distribution of difference between predicted time and the time of occurred earthquakes for 6, 12, 18,..., 42 months (the illustration of realibility by comparison between time difference distributions for all occured and predicted earthquakes with the time of tidal extremums).*

In Fig. 7, left the distributions of the difference between the times of predicted events, which occurred, calculated for 6, 12, 18,…, 42 months (starting from January, 2002), are presented. The distribution growth, without widening and its approximation to the Gaussian distribution with the time, is an argument for the causality-consequences origin between the correlation geomagnetic quake - tidal potential extremum and occurred earthquake. The number of earthquakes in Fig. 7, left is greater than the number of the predictions for the events. The explanation of this is that some earthquakes with greater magnitude are following by aftershocks. In Fig. 7, right, are presented the same distributions, but for the all occurred in the region (500 km) earthquakes. The difference illustrates the answer of the statement for "100% true predictions", which is based on the distances dependence of the signal for geomagnetic quake.

The fact, that the distributions of time differences between the tidal exstremum and all earthquakes, occurred in the same Tidal period, are flatter than the distributions of predicted earthquakes, can be consider as evidence that the correlation between geomagnetic quake and tidal extremum in the framework of physical mean of function $S_{ChtM}$ is a reliable precursor-Fig.8.

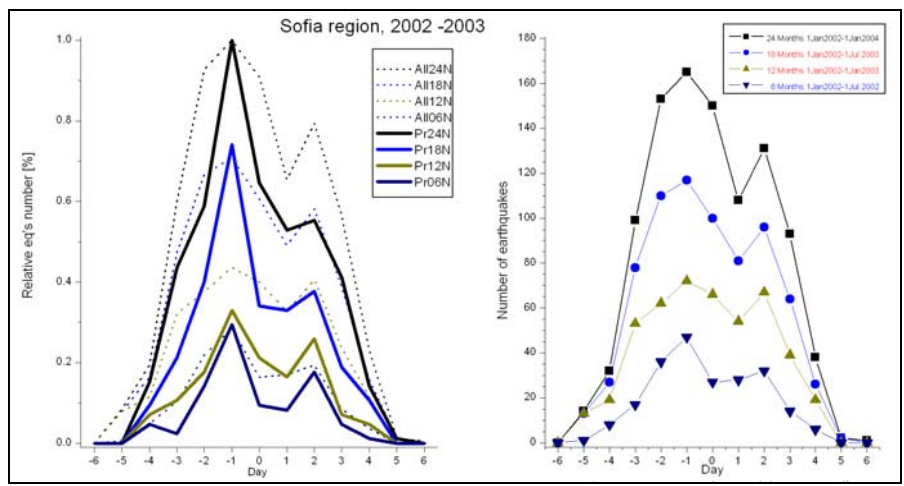

*Fig.8. Left: the comparison of the distributions of the difference between tidal extremums' time and the time of the predicted (solid curves) and all (dotted) occurred earthquakes for 6, 12, 18 and 24 months. Right: the distributions of the difference between tidal extremums' time and the time of all occurred earthquakes for 6, 12, 18 and 24 months.*

Next, Fig. 9 expresses the obvious fact that the incoming earthquake with grater magnitude can be predicted at greater distances. Nowadays estimation is that bigger earthquakes (magnitude>5) could be predicted for distances till 500- 600 km.

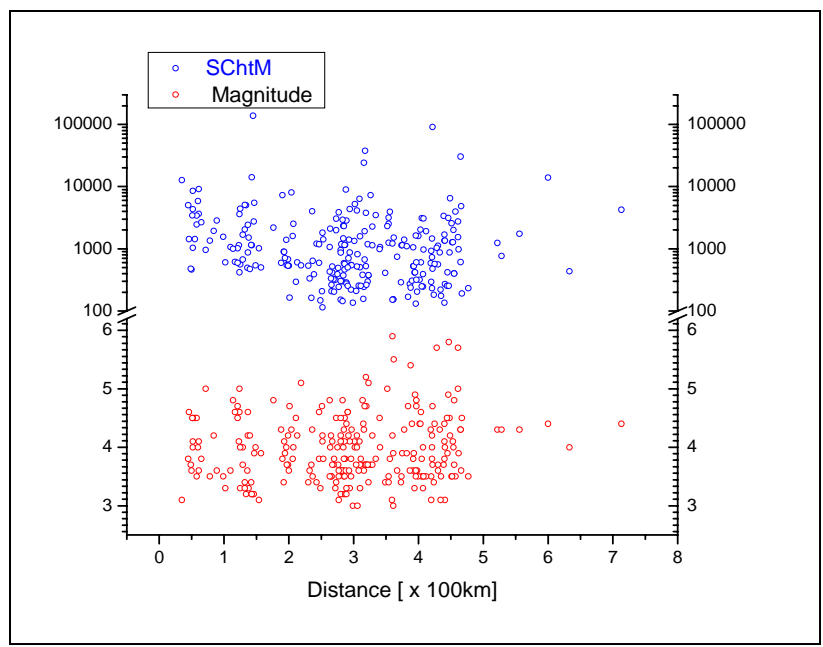

*Fig.9. The $S_{ChtM}$ function and magnitude for predicted and occurred earthquakes, Sofia region, 2002-2005.*

In Fig.10 the magnitude distribution of the predicted events is represented. It could be seen that the earthquakes with magnitude less then 3 are a small part (26 %) of all predicted and occurred earthquakes (Fig.11).

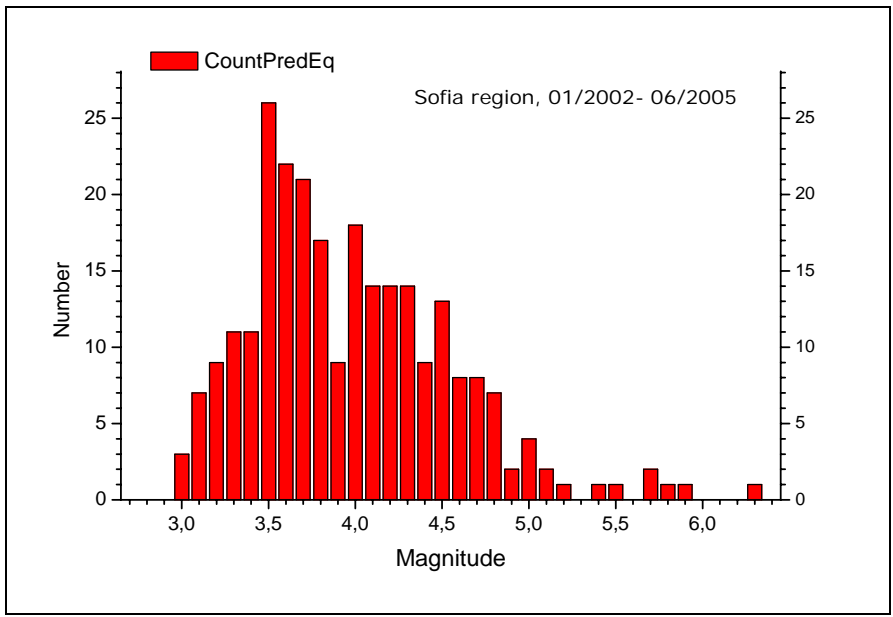

*Fig.10. The distribution of the magnitude for predicted earthquakes, Sofia region, 2002-2005.*

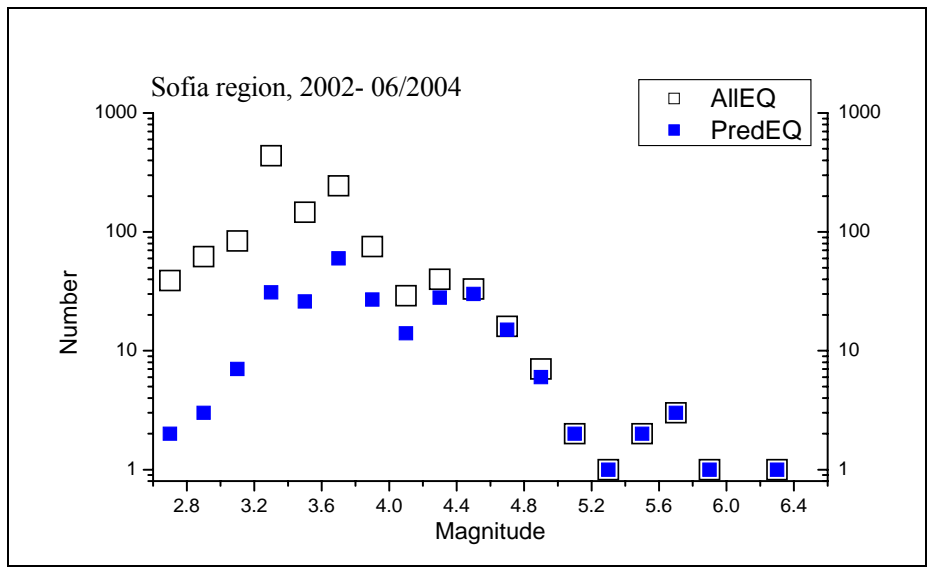

*Fig.11. The comparison of the number of all and predicted earthquakes*

*(it is seen, that the time of all occurred earthquakes with magnitude>4.9 are predicted)*

## 2.2. Skopje data

The Skopje triaxial fluxgate magnetometer is FGE model, Danish Meteorological Institute, with year sensors drift less then a few nT and temperature coefficient less or equal to 0.25 nT/C$^0$. The magnetometer was used in variometer mode with resolution 0.2 nT, 10 samples/sec. The signal SigD is daily averaged geometrical sum of normalized standard minute's deviations of components:

In the next Fig.12-17 are presented the analogs of Fig.5-11 for Skopje vector monitoring.

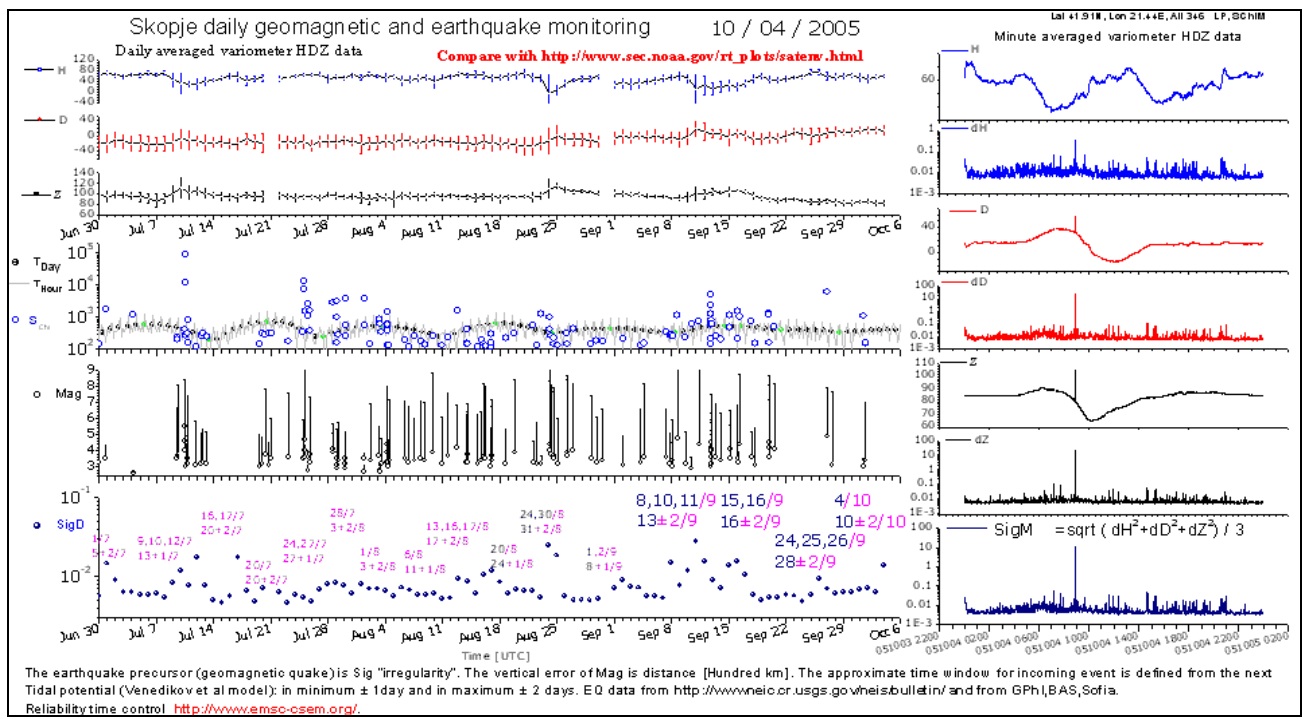

*Fig.12. The reliability of the time window prediction for incoming earthquake- June- October, 2005, Skopje.*

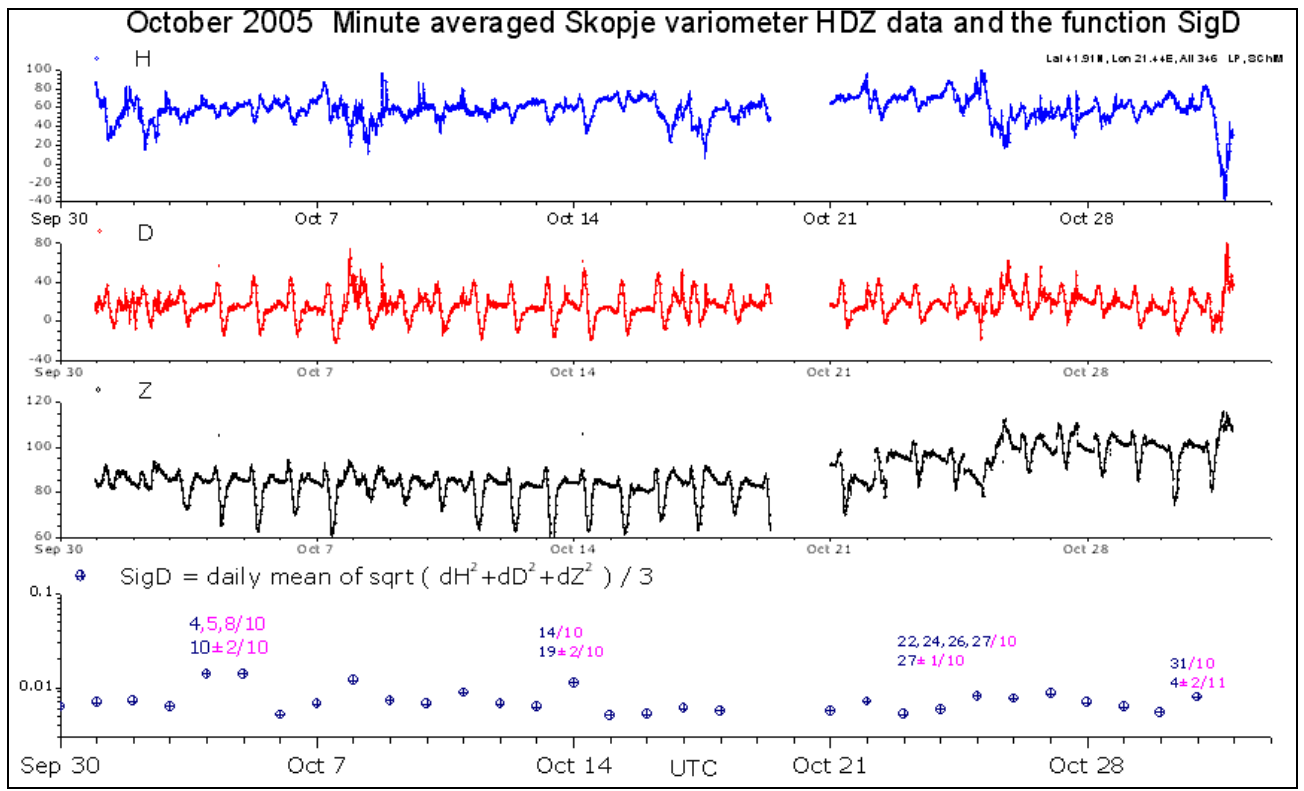

*Fig.13. The "monthly monitoring" figure- October, 2005, Skopje.*

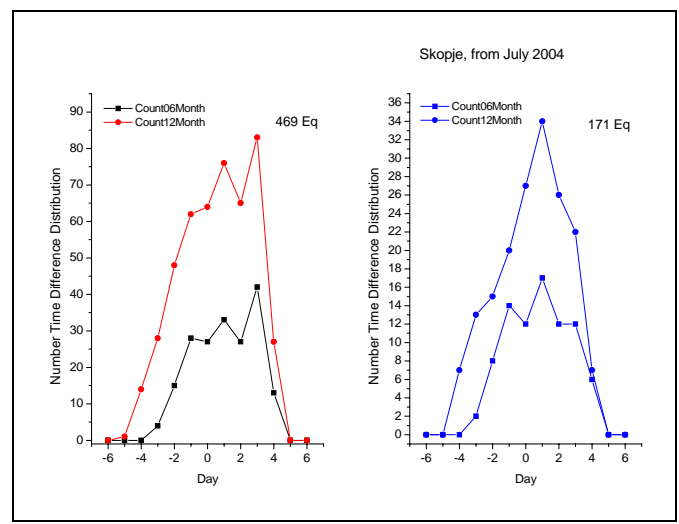

Fig14. The distribution of difference between predicted time and the time of occurred earthquakes for 6 and 12 months, Skopje (the time difference distribution for all (left) and predicted earthquakes)

In the next Fig. 15, 16 are presented the Skopje analogs of Fig. 8,9.

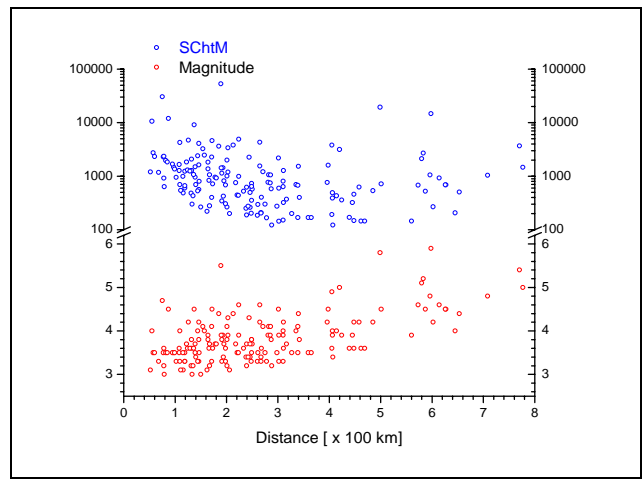

*Fig.15. The S$_{ChtM}$ function and magnitude for predicted and occurred earthquakes, Skopje region, 2004-2005.*

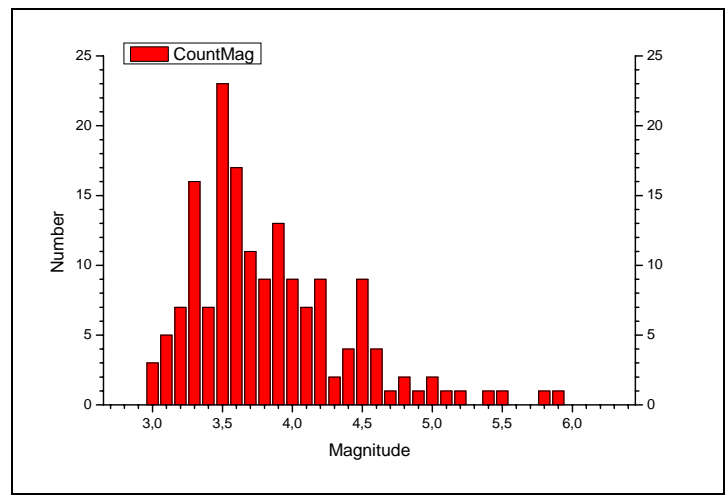

*Fig.16. The distribution of the magnitude for predicted earthquakes, Skopje, 2004-2005*

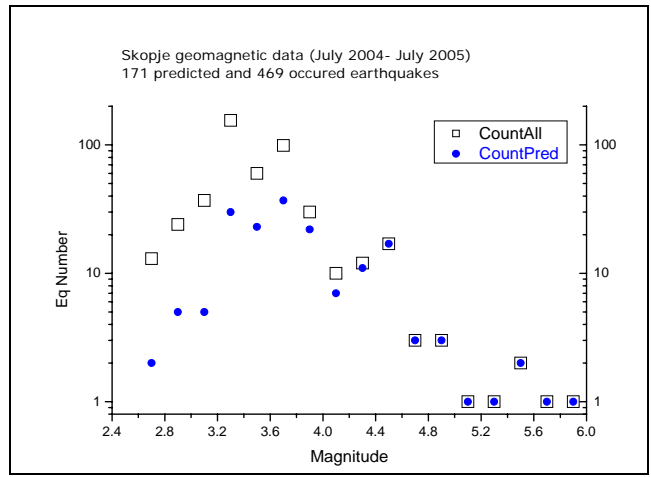

*Fig.17. The comparison of the number of all and predicted earthquakes*

*(it is seen, that the time of all occurred earthquakes with magnitude>4.5 are predicted)*

The independent control (in the framework of Strasbourg recommendations about publishing the earthquake predictions of the European Union for ethical and public security reasons) of the time window earthquake prediction reliability was organized in the framework of the Bulgarian Academy of Sciences, its Geophysical Institute and colleagues from collaboration EqTiPlaMagInt form Albania, Armenia, Georgia, Macedonia, Greece, Italy, Slovenia, who are interested in this topic of research.

## 2.2. The comparison of 2005 Skopje and Sofia geomagnetic monitoring and "when" earthquake predictions

In the Fig 18. are presented the distributions of the time difference between predicted and occurred earthquakes with magnitude greater then 3, SChtM greater then 200 for the region 500 km from Skopje and Sofia, correspondingly. The displacement of the Skopje peak from zero to one is connected with aftershocks and the high seismicity in Greece. The classification of the predicted earthquakes and their aftershocks is one nonsolved problem in our approach. After its solution will be possible to estimate more precisely the estimation of incoming time risk period- +/-1 day for the daily tidal minimum and +/-2 days for daily tidal maximum.

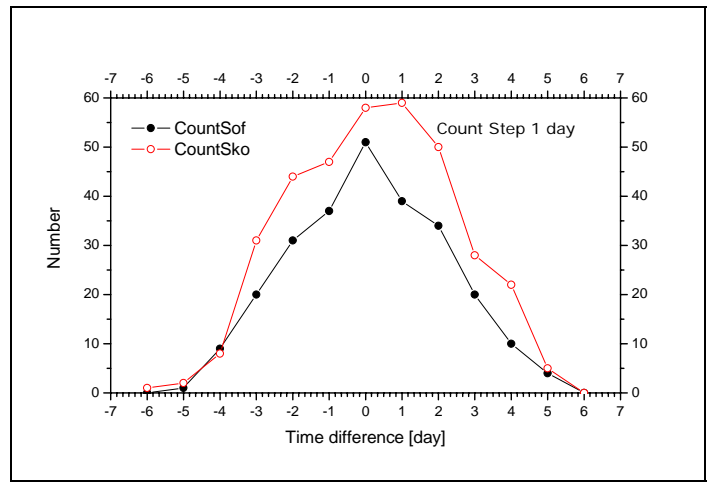

***Fig.18.*** *The distribution of time difference between predicted and occurred earthquakes in 2005 with magnutude greater then 3 in the 500 km radius from Sofia and Skopje, SChtM>200 and with count step equal to 1 day*

The distance sensibility of the approach as function of magnitude and distances is illustrated in the maps (Fig. 19,20, 21.) for earthquakes with magnitude greater then 5, 4 and 3 for the predicted earthquakes in the 500 km region. The reconstruction of the sensibility function is not performed because of indeterminations connected with the fact that the used magnetometers are different. The problem of its reconstruction will be possible after analyzing the geomagnetic monitoring data (at least 3 points!) for the "when, where" predicted earthquakes.

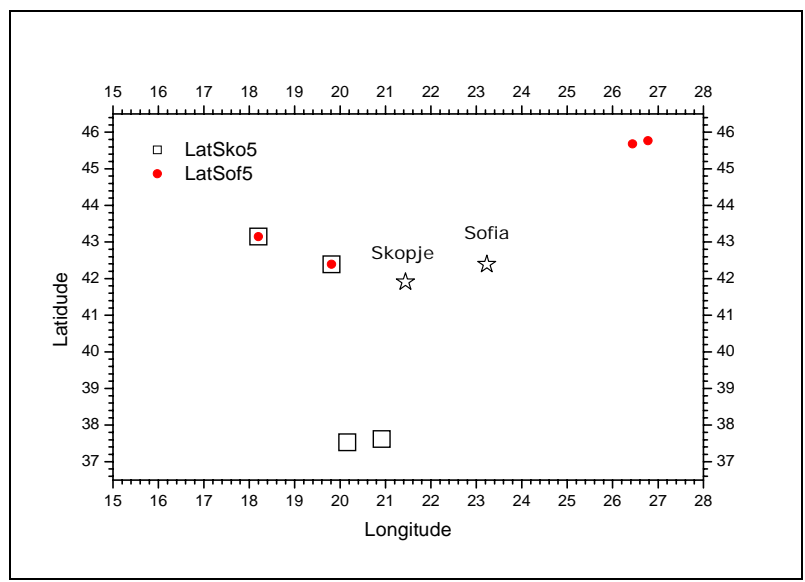

*Fig.19. The map of predicted from Skopje and Sofia earthquake with magnitude*

*greater then 5 in 500 km region from Skopje and Sofia for 2005*

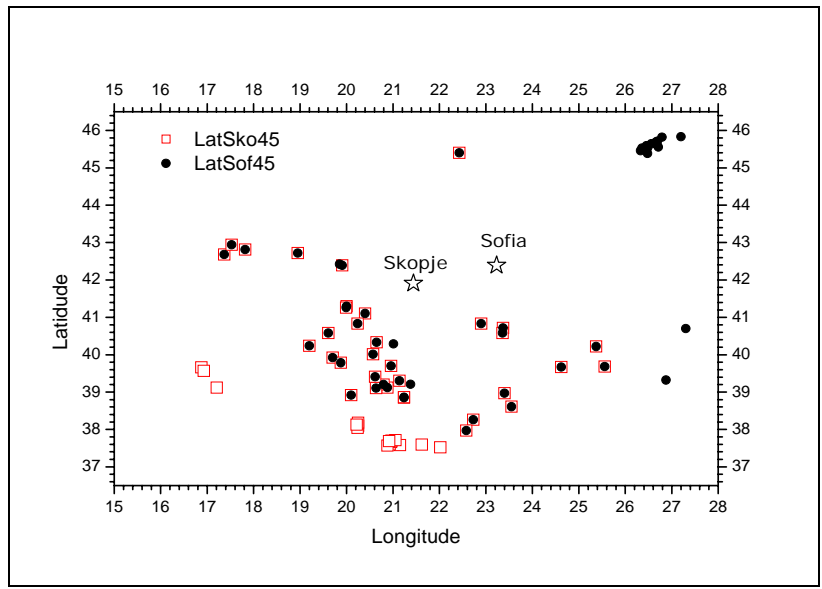

*Fig 20. The map of predicted from Skopje and Sofia earthquake with magnitude*

*greater then 4 and less, equal then 5 in 500 km region from Skopje and Sofia for  2005.*

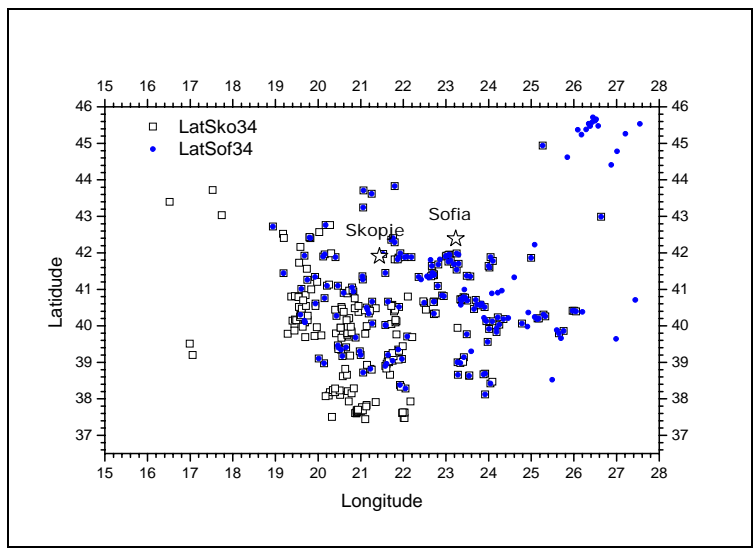

*Fig. 21. The map of predicted from Skopje and Sofia earthquake with magnitude*

*greater then 3 and less, equal then 4 in 500 km region from Skopje and Sofia for  2005.*

The confirmation of one geomagnetic component Sofia data from the vector Skopje results for reliability time window earthquake prediction can be consider as a first step for solution of "when, where and how" earthquake prediction at level "when". It is obvious that the occurred in the predicted time period earthquake with maximum value of function SChtM (proportional to the Richter energy density in the monitoring point) is the predicted earthquake. But some times there are more then one geomagnetic signals in one day or some in different days. It is not possible to perform unique interpretation and to choose the predicted earthquakes between some of them with less values of energy density. The solution of this problem will be the vector geomagnetic monitoring in at least 3 points, which will permit to start solving the inverse problem for estimation the coordinates of geomagnetic quake source as function of geomagnetic quake. The numbering of powers of freedom for estimation the epicenter, depth, magnitude and intensity (maximum values of accelerator vector and its dangerous frequencies) and the number of possible earthquake precursors show that the nonlinear system of inverse problem will be overdetermined.

Thus, the first prove, that in the framework of such complex approach, the "when, where and how" earthquake prediction problem can be solve will be the "when, where" prediction on the basis of at least 3 points for electromagnetic real time monitoring is essential. If the statistic estimation will be successful for a long enough period of time (6-12 months) and the established correlations are confirmed by the adequate physical model solutions, one could say that the earthquake prediction problem is under solving using the electromagnetic quake precursor.

Experimentally, the first attempts for estimation of the future epicenter can be perform on the basis of isolines distribution of geomagnetic quake and electropotential data set. Theoretically, the simplest inverse problem is to estimate the coordinates of vertical Earth surface – Ionosphere electrical current which generates the geomagnetic quake. Such time of modeling has to be perform step by step by including the volume of solid state phase shifts and frequency characteristics like function of stress, depth and geology.

The posteriori analysis for England, Alaska, India, Kamchatka, Hokkaido regions on the basis of second vector Intermagnet data [INTERMAGNET, WS], also confirm the reliability of geomagnetic quake as imminent earthquake precursors (Mavrodiev 2003 a, b, 2004).

### 3. Conclusion

The correlations between the local geomagnetic quake and incoming earthquakes, which occur in the time window defined from tidal minimum (± 1 day) or maximum (± 2 days) of the Earth tidal gravitational potential are tested statistically. The distribution of the time difference between predicted and occurred events is going to be Gaussian with the increasing of the statistics.

This result can be interpreted as first reliable approach for solving the "when" earthquakes prediction problem using the geomagnetic data.

*Acknowledgement:* Thanks to Cht. S. Mavrodiev for technical support. The financial support from Bulgarian Scientific Fund (contract BM-9/2005) is highly appreciated.

**References:**

Abercrombie, R. E., Earthquake source scaling relationships from -1 to 5 $M_L$, using seismograms recorded at 2.5 km depth, J. Geophys. Res., 100, 24015-24036, 1995a

Abercrombie, R. E., Earthquake locations using single-station deep borehole recordings: implications for microseismicity on the San Andreas fault in southern California, J. Geophys. Res., 100, 24003-24014, 1995b

Aki K., Earthquake prediction, societal implications, U.S. National Report to IUGG, 1991-1994, Rev. Geo-phys. Vol. 33 Suppl., © 1995 American Geophysical Union, *http://www.agu.org/revgeophys/aki00/aki00.html* , 1995.

Bragin, Y. A., Bragin O.A., Bragin V.Y., Reliability of Forecast and Lunar Hypothesis of Earthquakes, Report at XXII General Assembly of the International Union of Geodesy and Geophysics (IUGG), Birmingham, UK, July 18 – 30, 1999.

Balch C., Geomagnetic Indices: Real-Time Production and Forecasting, IUGG 2003 General Assembly, Sapporo, Japan, June 30-July 11, 2003, http://www.sel.noaa.gov/info/Kindex.html, 2003.

Contadakis M., Biagi P., Zschau J., 27th General Assembly, Acropolis, Nice, France, Natural Hazards, NH10, Seismic hazard evaluation, precursory pheno-mena and reliability of prediction, http://www.cosis.net/members/frame.php?url=www.copernicus.org/EGS/EGS.html, April 2002.

Dean R., www.earthquakeforecast.org, 2003.

Dierks O., J. Neumeyer, Comparison of Earth Tides Analysis Programs, Geo Forschungs Zentrum Potsdam, Division 1

 Freund F.T., A.Takeuchi, B. W.S. Lau, Electric Currents Streaming out of Stressed Igneous Rocks – A Step Towards Understanding Pre-Earthquake Low Frequency EM Emissions, Special Issue "Recent Progress in Seismo Electromagnetics", Guest Editors M. Hayakawa, S. Pulinets, M. Parrot, and O. A. Molchanov, Phys. Chem. Earth, 2006

Telegrafenberg, 14473 Potsdam, Germany,http://www.ksb.be/ICET/bim/text/dierks.htm, 2002

Dudkin F., De Santis A., Korepanov V., Active EM sounding for early warning of earthquakes and volcanic eruptions, *Physics of the Earth and Planetary Interiors,* Vol. 139, Issues 3-4, pp. 187-195 (2003)

Eftaxias K., P.Kapiris, J.Polygiannakis, N.Bogris, J.Kopanas, G.Antonopoulos, A.Peratzakis and V.Had-jicontis, Signature of pending earthquake from electro-magnetic anomalies, Geophysical Research Letters, Vol.28, No.17, pp.3321-3324, September 2001.

Eftaxias K., P.Kapiris, E.Dologlou, J.Kopanas, N.Bogris, G.Antonopoulos, A.Peratzakis and V.Hadjicontis, EM anomalies before the Kozani earthquake: A study of their behaviour through laboratory experiments, Geophysical Research Letters, Vol. 29, No.8 10.1029/ 2001 GL013786, 2002.

Geller R.J., Jackson D. D., Kagan Y. Y., Mulargia F., Earthquakes cannot be predicted, *SCIENCE,* 14 March 1997; 275: 1616-0,*http://scec.ess.ucla.edu/%7Eykagan/perspective.html, 1997.*

Geller, R. J., Debate on evaluation of the VAN method: Editor's introduction, Geophys. Res. Lett., 23(11), 1291-1294, 10.1029/96GL00742, 1996.

Geomagnetic data, Intermagnet, http://www.intermagnet.org/Data_e.html, ttp://geomag.usgs.gov/frames/mag_obs.htm, http://swdcwww.kugi.kyoto-u.ac.jp/wdc/Sec3.html, http://www.intermagnet.org/Magobs_e.html, from 1986.

Giardini D., Jiménez M.-J, Gruntha G., The ESC- SESAME Unified Hazard Model for the European-Mediterranean Region, EMSC/CSEM, Newsletter, 19, 2-4., 2003.

Hayakawa, M. and Fujinawa, Y. (Eds), Electroma-gnetic Phenomena Related to Earthquake Prediction, Terrapub, Tokyo, 677 pp. 1994.

Hayakawa, M., Fujinawa, Y., Evison, F. F., Shapiro, V. A., Varotsos, P., Fraser-Smith, A. C., Molchanov, O. A., Pokhotelov, O. A., Enomoto, Y., and Schloessin, H. H., What is the future direction of investigation on electromagnetic phenomena related toearthquake prediction?, in: Electromagnetic phenomena related to earthquake prediction, (Eds) Hayakawa, M. and Fujinawa, Y.,Terrapub, Tokyo, 667–677, 1994.

Hayakawa, M. (Ed): Atmospheric and Ionospheric Electromagnetic Phenomena Associated with Earthquakes, Terrapub, Tokyo, 996 pp., 1999.

Hayakawa, M., Ito, T., and Smirnova, N., Fractal analysis of ULF geomagnetic data associated with the Guam eartquake on 8 August 1993, Geophys, Res. Lett., 26, 2797–2800, 1999.

Hayakawa, M., Itoh, T., Hattori, K., and Yumoto, K., ULF electromagnetic precursors for an earthquake at Biak, Indonesia oin 17 February 1996, Geophys, Res. Lett., 27, 10, 1531–1534, 2000.

Hayakawa, M. and Molchanov, O. (Eds): Seismo Electromagnetics Lithosphere-Atmosphere-Ionosphere coupling, Terrapub, Tokyo, 477 pp., 2002.

Hrvoic Ivan, http://www.gemsys.ca/contact_contact.htm

Kanamori, H. and E. E. Brodsky, The Physics of Earthquakes, *Reports on Progress in Physics*, **67**, 1429 - 1496, 2004.

Karakelian, D., Klemperer, S.L., Fraser-Smith, A.C., and Beroza, G.C., A Transportable System for Monitoring Ultralow Frequency Electromagnetic Signals Associated with Earthquaes, *Seismo-logicalResearch Letters*, 71, 4, 423-436, 2000.

Karakelian, D., Klemperer, S.L., Thompson, G.A., and Fraser-Smith, A.C. In review. *Proc. 3rd conf. "Tectonic problems of the San Andreas Fault", Stanford, CA, 2001.*

Karakelian, D., Beroza, G.C., Klemperer, S.L.,, and Fraser-Smith, A.C. Analysis of ultra-low frequency electromagnetic field measurements associated with the 1999 M 7.1 Hector Mine earthquake sequence. *Bull. Seismolog. Soc. Am., v. 92,* pp. 1513-1524, 2002

Keilis-Borok V. I., http://www.mitp.ru, from 2000.

Klemperer S., T. Fraser-Smith, G. Beroza, D. Karakelian, Stanford University, Ultra-low frequency electromagnetic monitoring within PBO, http://www.scec.org/news/00news/images/pbominiproposals/Klempererpbo40.pdf, 2003.

Knopoff, L., Earth tides as a triggering mechanism for earthquakes, Bull. Seism. Soc. Am., Vol. 54, pp. 1865 – 1870, 1964.

Knopoff L., Earthquake prediction: The scientific challenge, and other Papers from an NAS Colloquium on Earthquake Prediction, http://www.pnas.org/content/vol93/issue9/#COLLOQUIUM, 1996.

Korepanov V., Dudkin F., Electromagnetic precursors of seismic catastrophes study, *Progress in electromagnetics research symposium (PIERS 2003).* Proceedings, Singapore, p. 198, 2003.

Korepanov V., Klymovych Ye., Rakhlin L., Electromagnetic instrumentation – new developments for geophysics, Int. Geophysical Conf., Moscow, Russia, September 2003

Larkina V.I., Ruzhin Yu.Ya, Wave Satellite Monitoring of Earthquake Precursors in the Earth Plasmasphere, in 1[st] International Workshop on Earthquake Prediction, Subcommission on Earth-quake Prediction Studies (SCE) of the European Seismological Commission scheduled, *Athens, Greece, November 2003*, http://www.gein.noa.gr/ services/ Workshop.htm, 2003.

Ludwin, R.S., 2001, Earthquake Prediction, Washing-ton Geology, Vol. 28, No. 3, May 2001, p. 27, 2001.

Main I., Is the reliable prediction of individual earth-quakes a realistic scientific goal?, Debate in *NATURE, http://www.nature.com/nature/debates/earthquake/equake_contents.html, 1999a*

Main I., Earthquake prediction: concluding remarks. Nature debates, Week 7, 1999b.

Mavrodiev S.Cht., Applied Ecology of the Black Sea, ISBN 1-56072- 613- X, 207 Pages, Nova Science Publishers, Inc., Commack, New York 11725, 1998.

Mavrodiev S.Cht., Thanassoulas C., Possible correlation between electromagnetic earth fields and future earthquakes, INRNE-BAS, Seminar proceedings, 23- 27 July 2001, Sofia, Bulgaria, ISBN 954-9820-05-X, 2001, http://arXiv.org/abs/physics/0110012, 2001.

Mavrodiev S.Cht., The electromagnetic fields under, on and up Earth surface as earthquakes precursor in the Balkans and Black Sea regions, http://arXiv.org/ abs/ 0202031, February, 2002a.

Mavrodiev S.Cht., On the short time prediction of earth-quakes in Balkan- Black Sea region based on geomagnetic field measurements and tide gravitational potential behavior, http://arXiv.org/abs/physics/0210080, Oct, 2002b.

Mavrodiev S.Cht., The electromagnetic fields under, on and over Earth surface as "when, where and how" earthquake precursor, Workshop on Gujarat Earthquake, Kanpur, India, January 2003,

http://www.arxiv.org/ftp/physics/papers/0302/0302033.pdf, January 2003a.

Mavrodiev S.Cht., The electromagnetic fields under, on and over Earth surface as "when, where and how" earthquake precursor, European Geophysical Society, Geophysical Research Abstracts, Vol.5, 04049, 2003b.

Mavrodiev S.Cht., On the Reliability of the Geomagnetic Quake Approach as Short Time Earthquake's Precursor for Sofia Region, Natural Hazards and Earth System Science, Vol. 4, pp 433-447, 21-6-2004

Molher A.S., Earthquake / earth tide correlation and other features of the Susanville, California, earthquake sequence of June-July 1976., Bull. Seism. Soc.Am., Vol. 70, pp. 1583 – 1593, 1980.

Earthquake data, NEIC, http://wwwneic.cr.usgs.gov/neis/ bulletin/,http://www.emsc-csem.org/, http://www.iris.edu/ quakes/eventsrch.htm, http://www.geophys.washington.edu/ seismosurfing.html, from 1972.

NOAA, http://www.sec.noaa.gov/SWN/, http://www.sec.noaa.gov/rt_plots/satenv.html, http://www.sec.noaa.gov/ rt_plots/xray_5m.html, from 1972.

Oike, K. and Ogawa, T., Observations of electromag-netic radiation related with the occurrence of earthquake, Annu. Rep. Desaster, Prev. Res. Inst. Kyoto Univ., 25, B-1, 89–100, 1982.

Oike, K. and Yamada, T., Relationship between shallow earthquakes and electromagnetic noises in the LF and VLF ranges, in: Electromagnetic phenomena related to earthquake prediction, (Eds) Hayakawa, M. and Fujinawa, Y., Terrapub, Tokyo, 115–130, 1994.

Pakiser L, Shedlock K.M., Predicting earthquakes, USGS, http://earthquake.usgs.gov/hazards/prediction.html, 1995.

Papadopoulos G., 1st International Workshop on Earth-quake Prediction, Subcommission on Earthquake Prediction Studies (SCE) of the European Seismological Commission scheduled, *Athens, Greece, November 2003*, http://www.gein.noa.gr/services/Workshop.htm, 2003.

Qian, S., Yian, J., Cao, H., Shi, S., Lu, Z., Li, J., and Ren, K., Results of observations on seismo-electro-magnetic waves at two earthquake experimental areas in China, in: Electromagnetic phenomena related to earthquake prediction, (Eds) Hayakawa, M. and Fujinawa, Y., Terrapub, Tokyo, 205–211, 1994.

Panza G. F., Romanelli F., Vaccari F., SEISMIC WAVE PROPAGATION IN LATERALLY HETEROGENEOUS ANELASTIC MEDIA: THEORY AND APPLICATIONS TO SEISMIC ZONATION, Advances in Geophysics, Vol. 43, 1-95, 2000.

Ramesh P. Singh, Editor, Monitoring of Changes Related to Natural and Manmade Hazards Using Space Technology, Advances in Space Research, Volume 33, Issue 3, 2004, Page 243

Ryabl, A., Van Wormer, J. D., Jones, A. E., Triggering of micro earth-quakes by earth tides and other features of the Truckee, California, earthquake sequence of September 1966, Bull. Seism. Soc. Am., Vol. 58, pp. 215 – 248, 1968.

Saraev A.K., Pertel M.I., Malkin Z.M., Correction of the electromagnetic monitoring data for tidal variations of apparent resistivity, Journal of applied geophysics, 49, p.91-100, 2002.

Shlien S., Earthquake - tide correlation, Geophys. J. R. Astr. Soc., Vol. 28, pp. 27– 34, 1972.

Shirley J., Lunar and Solar periodicities of large earthquakes: Southern California and the Alaska Aleutian Islands seismic region., Geophysical Journal, Vol. 92., pp. 403 – 420, 1988.

Silina A.S., Liperovskaya E.V., Liperovsky1 V.A., Meister C.-V, Ionospheric phenomena before strong earthquakes, Natural Hazards and Earth System Sciences, 1: 113–118, 2001.

Sounau M., Sounau A., Gagnepain J., Modeling and detecting interaction between earth tides and earthquakes with application to an aftershock sequence in the Pyrenees, Bull. Seism. Soc. Am., Vol. 72, pp. 165 – 180, 1982.

St-Laurent F., J. S. Derr, Freund F. T. , Earthquake Lights and the Stress-Activation of Positive Hole Charge Carriers in Rocks, Special Issue "Recent Progress in Seismo Electromagnetics", Guest Editors M. Hayakawa, S. Pulinets, M. Parrot, and O. A. Molchanov, Phys. Chem. Earth, 2006

Suhadolc, P., Fault-plane solutions and seismicity around the EGT southern segment. In: R. Freeman and St. Müller (Editors), Sixth EGT Workshop: Data Compilations and Synoptic Interpretation, European Science Foundation,

Strasbourg, pp. 371-382, 1990a.

Suhadolc, P., Panza, G.F., Marson, I., Costa, G. and Vaccari, F., 1992. Analisi della sismicità e meccanismi focali nell'area italiana. Atti del Convegno del Gruppo Nazionale per la Difesa dai Terremoti, Pisa, **1**, 157-168, 1990b.

Tamrazyan D.P., Tide- Forming Forces and Earthquakes, ICARUS, Vol.7, pp.59-65, 1967.

Tamrazyan D.P., Principal Regularities in the Distribution of Major Earthquakes Relative to Solar and Lunar Tides and Other Cosmic Forces, ICARUS, Vol.9, pp.574-592, 1968.

Tamura Y., Sato T., Ooe M., Ishiguro M., A *procedure for tidal analysis with a Bayesian information criterion*, Geophysical Journal International, No. 104, pp. 507-516, Oxford: 1991.

Thanassoulas, C., Determination of the epicentral area of three earthquakes (Ms>6R) in Greece, based on electrotelluric currents recorded by the VAN network., Acta Geophysica Polonica, Vol. XXXIX, no. 4, 373-387, 1991.

Thanassoulas C., Earthquake prediction based on electrical signals recorded on ground surface- An integrated methodology answering on "where", "when" and "of what magnitude" a large EQ will occur,http://www.earthquakeprediction.gr/,from 1999.

Thanassoulas, C., Tsatsaragos, J., Klentos, V., Deter-mination of the most probable time of occurrence of a large earthquake., Open File Report A. 4338, IGME, Athens, Greece, 2001a.

Thanassoulas, C., Klentos, V., Very short-term (+/- 1 day, +/- 1 hour) time-prediction of a large imminent earthquake, The second paper, Institute of Geology and Mineral Exploration (IGME), Athens, Greece, Open File Report A. 4382, pp 1-24, 2001b.

Thanassoulas, C., Klentos, V., The "energy-flow mo-del" of the earth's lithosphere. Its application on the prediction of the "magnitude" of an imminent large earthquake, The third paper, Institute of Geology and Mineral Exploration (IGME), Athens, Greece, Open file report: A.4384, pp.1-20, 2001c.

Thanassoulas, C., and Klentos,V., The "energy-flow mo-del" of the earth's lithosphere, its application on the prediction of the "magnitude" of an imminent large Earthquake, The "third paper", Institute of Geology and Mineral Exploration (IGME), Athens, Greece, Open File Report A. 4384, pp 1-20, 2001d.

Tsatsaragos J., Predicting earthquakes, (with Thanassoulas, C. till 2002), http://users.otenet.gr/~bm-ohexwb/alert2.htm, from 1999.

USGS, http://quake.wr.usgs.gov/research/index.html

[USGS Pf], The Parkfield Experiment- Capturing What Happens in an Earthquake, http://geopubs.wr.usgs.gov/fact-sheet/fs049-02/, 2002.

Ustundag B., Earthquake Forecast, Data Acquisition and Evaluation Page, http://www.deprem.cs.itu.edu.tr/, from 2001.

Varotsos, P. and Alexopoulos, K., Physical properties of the variations of the electric field of the Earth preceding earthquakes, I, Tectonophysics, 110, 93–98, 1984a.

Varotsos, P. and Alexopoulos, K., Physical properties of the variations of the electric field of the Earth preceding earthquakes, II, Tectonophysics, 110, 99–125, 1984b.

Varotsos, P., Lazaridou, M., Eftaxias, K., Antonopoulos, G., Makris, J., and Kopanas, J., Short-term Earthquake Prediction in Greece by Seismic Electric Signals, in: A Critical Review of VAN: Earthquake prediction from Seismic Electric Signals, (Ed) Ligthhill, Sir J., World Scientific Publishing Co., Singapore, 29– 76, 1996.

Venedikov A., Arnoso R., Program VAV/2000 for Tidal Analysis of Unevenly Spaced Data with Irregular Drift and Colored Noise, J. Geodetic Society of Japan, vol.47, 1, 281- 286, 2001.

Venedikov A.P., Arnoso R., Vieira R., A program for tidal data processing, Computers & Geosciences, vol. 29, no.4, pp. 487-502, 2003.

Wenzel, H.-G., The nanogal software: Earth tide data processing package ETERNA 3.30. In: Marees Terrestres Bulletin d'Informations, No. 124, pp. 9425-9439, Paul Melchior (Edit.). Bruxelles: Association Internationale de Géodésie, 1996a.

Wenzel, H.-G., *ETERNA Manual*. Version 3.30, Karl-sruhe, Black Forest Observatory, 1996b.

Zhonghao Shou, Earthquake Clouds and Short Term Prediction, http://quake.exit.com/, from 1999.

Zhonghao Shou, D. Harrington, Bam Earthquake Prediction & Space Technology, Earthquake Prediction Center, *500E 63rd 19K, New York, NY 10021, 2005.*